\documentclass[aps,showpacs,preprintnumbers,twocolumn,amsmath,amssymb,prl,floatfix,superscriptaddress]{revtex4} 

\usepackage{graphicx}

\usepackage{ifpdf}

\ifpdf
\usepackage{epstopdf}
\usepackage[pdftex]{hyperref}
\else
\usepackage[hypertex,ps2pdf,dvips,colorlinks,urlcolor=blue,citecolor=blue]{hyperref}
\fi

\newcommand\be{\begin{equation}}
\newcommand\bes{\begin{subequations}}
\newcommand\esu{\end{subequations}}
\newcommand\ee{\end{equation}}
\newcommand\erf[1]        {\eqref{#1}}
\newcommand{\ud}          {\mathrm d}

\newcommand\p             {\partial}
\newcommand\psid          {\psi^{\dagger}}

\newcommand\lam             {\lambda}
\newcommand\ga             {\gamma}

\newcommand\rholl          {\rho_{\text{LL}}}
\newcommand\vev[1]{{\langle#1\rangle}}

\newcommand\doi[2]        {\href{http://dx.doi.org/#1}{#2}}

\begin{document}

\title{Interaction quenches in the 1D Bose gas}

\author{M\'arton Kormos}
\affiliation{Department of Physics and Astronomy, Rice University, Houston, Texas 77005, USA}
\affiliation{Dipartimento di Fisica dell'Universit\`a di Pisa and INFN, 56127 Pisa, Italy}
\author{Aditya Shashi}
\affiliation{Department of Physics and Astronomy, Rice University, Houston, Texas 77005, USA}
\affiliation{Department of Physics, Harvard University, Cambridge, MA 02138, USA}
\author{Yang-Zhi Chou}
\affiliation{Department of Physics and Astronomy, Rice University, Houston, Texas 77005, USA}
\author{Jean-S\'ebastien Caux}
\affiliation{Institute for Theoretical Physics, University of Amsterdam, Science Park 904, Postbus 94485, 1090 GL Amsterdam, The Netherlands}
\author{Adilet Imambekov}
\affiliation{Department of Physics and Astronomy, Rice University, Houston, Texas 77005, USA}

\date{\today}

\begin{abstract}

The non-equilibrium dynamics of integrable systems are special: there is substantial evidence that after a quantum quench they do not thermalize but their asymptotic steady state can be described by a Generalized Gibbs Ensemble (GGE). Most of the studies on the GGE so far have focused on models that can be mapped to quadratic systems while analytic treatment in non-quadratic systems remained elusive. We obtain results on interaction quenches in a non-quadratic continuum system, the 1D Bose gas described by the integrable Lieb--Liniger model. We compute local correlators for a non-interacting initial state and arbitrary final interactions as well as two-point functions for quenches to the Tonks--Girardeau regime. We show that in the long time limit integrability leads to significant deviations from the predictions of the grand canonical ensemble.

\end{abstract}


\maketitle

Whether and how an isolated quantum system equilibrates or thermalizes
are fundamental questions in understanding non-equilibrium
dynamics. The answers can also shed light on the applicability of
quantum statistical mechanics to closed systems. While these questions
are very hard to study experimentally in the condensed matter setup,
they have become accessible in ultracold quantum gases due to 
recent experimental advances \cite{review}. Thanks to their
unprecedented tunability, ultracold atomic systems allow for the
study of non-equilibrium quantum dynamics of almost perfectly isolated
strongly correlated many-body systems in a controlled way. These
experiments \cite{kinoshita,greiner,hofferberth,trotzky,cheneau,gring,haller,schneider,ronzheimer} triggered a revival of theoretical studies on issues of thermalization \cite{rev,rigorous,cardycalabrese,quench,LLquench,ising,luttinger,GGE,gge_various,gge_non2,fioretto,mossel,local,js-fabian,truncGGE}.
The list of fundamental questions include whether stationary values of local
correlation functions are reached in a system brought out of
equilibrium, and if so, how they can be characterized. Can
conventional statistical ensembles describe the state? Is there any
kind of universality in the steady state and the way it is approached?

The absence of thermalization of a 1D bosonic gas reported in Ref.\ 
\cite{kinoshita} brought to light the special role of
integrability. The observed lack of thermalization was attributed to the fact
that the system was very close to an integrable one, the Lieb--Liniger
(LL) model \cite{LL} which is the subject of our Letter. The dynamics
of integrable systems are highly constrained by the presence of a large
number of conserved charges \cite{sutherland} in addition to the total particle
number, momentum, and energy, thus they are not expected
to thermalize. The so-called Generalized Gibbs
Ensemble (GGE) was proposed \cite{GGE} to capture the long-time
behavior of integrable systems brought out of equilibrium. 
The density matrix is
\be \hat\rho_{\text{GGE}}=\frac{e^{-\sum_m \beta_m\hat Q_m}}{Z_\text{GGE}}\,,
\label{eq:GGErho}
\ee
where the generalized ``chemical potentials"
$\{\beta_m\}$ are fixed by the expectation values $\vev{\hat Q_m}$ in the initial state, and
$Z_\text{GGE}=\mathrm{Tr}\left[e^{-\sum_m \beta_m\hat Q_m}\right]$.
The GGE proposal was tested successfully by various numerical and analytic approaches \cite{gge_various,gge_non2,fioretto,mossel}. 

Recently, {\em locality} has emerged as a crucial ingredient in the  understanding of equilibration and the meaning of a steady state \cite{local}. While the whole system starting in an initial pure state clearly cannot evolve into a mixed state, its subsystems are fully described by a reduced density matrix obtained by tracing out the rest of the system that acts as a bath for the subsystem. There is substantial evidence that this density matrix is thermal for generic systems and given by the GGE for integrable systems. 
%
%
Naturally, it is the {\em local} conserved charges that are to be used in the GGE density matrix. 

The GGE was studied mostly in models which can be
mapped to quadratic bosonic or fermionic systems 
where the conserved charges are given by the mode occupation
numbers. While some of these models are paradigmatic, like the Ising
or Luttinger models, a prominent class of non-trivial integrable
systems has not been sufficiently explored, namely those solvable by
the Bethe Ansatz (BA). In these models, the local conserved charge operators are usually known
but cannot be expressed as mode occupations. 

The work \cite{fioretto} focused on integrable quantum field theories 
and demonstrated that the long-time limit of expectation values are given by a GGE, assuming a special initial state.
In Ref.~\cite{js-fabian} it was shown for BA integrable models that in the thermodynamic limit the time evolution of local observables after a quantum quench is captured by a saddle point state, and their $t\to\infty$ asymptotic values are given by their expectation values in this state.  The saddle point state can be determined using the expectation values of the charges in the initial state.

In this Letter, we focus on a BA solvable {\em continuum} model: we derive experimentally testable predictions for the long time behavior of the LL model after an interaction
quench~\cite{LLquench} combining Bethe
Ansatz methods and GGE. For a non-interacting initial state and
arbitrary final interactions we calculate expectation values of point-localized operators, while for quenches to the fermionized Tonks--Girardeau regime we obtain exact results on two-point correlation functions.

\paragraph*{The model.---} The LL model describes a system of identical
bosons in 1D interacting via a Dirac-delta potential.
The Hamiltonian is given by \cite{LL}
\be
\hat H=-\sum_i^N\frac{\p^2}{\p x_i^2}+2c\sum_{i<j}\,\delta(x_i-x_j)\,,
\ee
which in the second quantized formulation takes the form
\be
\hat H= \int_0^L\ud x\,\left(\p_x\hat\psi^\dagger\p_x\hat\psi+
c\,\hat\psi^\dagger\hat\psi^\dagger\hat\psi\hat\psi\right)\,,
\label{H2}
\ee
where $c>0$ in the repulsive regime we wish to study, and 
for brevity we set $\hbar=1$ and the boson mass to be equal to
$1/2$. The dimensionless coupling constant is given by
$\ga=c/n$, where $n=N/L$ is the density of the gas. In cold atom
experiments $\ga$ is a function of the 3D scattering length and the 1D
confinement \cite{olshanii}. The exact spectrum and thermodynamics of
the model can be obtained via Bethe Ansatz \cite{LL,KBI}. The many-body eigenfunctions $\phi(\{x_i\})$ of $\hat H$ satisfy the boundary condition
\be
\left.\left(\frac{\p}{\p x_j}-\frac{\p}{\p x_k}-c\right)\phi(x_1,\dots,x_N)\right|_{x_j=x_k+0}=0\,,
\label{cusp}
\ee
whenever the coordinates of two particles coincide, thus the wave functions have cusps. 
The eigenstates on a ring can be expressed in terms of $N$ quasimomenta $\{\lam_j\}$ that satisfy a set of algebraic equations, the Bethe equations.
The eigenvalues of the mutually commuting {\em local} conserved charges can be computed as $\vev{\hat Q_m}= \sum_j \lam_j^m,$ in particular, the energy is simply $E=\vev{\hat Q_2}=\sum_j\lam_j^2$ \cite{KBI}. 
In the thermodynamic limit (TDL), a mixed state
is captured by a continuous density of quasimomenta, $\rholl(\lam)$\cite{yang}. All quasimomenta are coupled to each other by the Bethe equations and thus $\rholl(\lam)$ as well as the density of ``holes'' satisfies integral equations, the Thermodynamic Bethe Ansatz (TBA) equations.
This approach was developed for thermal equilibrium but it can be generalized to the case of the GGE \cite{mossel}.

\paragraph*{Divergence of the local conserved charges.---}

The simplest way to bring a system out of equilibrium is a sudden
change of one of its parameters, a quantum quench \cite{cardycalabrese}. In a cold atom
setting such a quench could be achieved by a rapid change of the
transverse confinement or the scattering length.  We will compute the
predictions of the GGE for a sudden quench of the interaction
parameter $c$ starting from the ground state of the $c=0$ system, a pure non-interacting BEC (although we expect our results to be also valid for small initial interactions) and compare them with those of the grand canonical ensemble (GCE).

In order to describe the final state in terms of the distribution $\rholl(\lam)$, one needs to find the
expectation values of the conserved charges $\hat Q_m$ right after the
quench, i.e. in the BEC-like ground state of free bosons.  
The density $\rholl(\lam)$ is then found, in principle, by solving the problem of
moments defined by $\vev{\hat Q_m}=L\int\ud\lam\,\rholl(\lam)\,\lam^m.$ 
The first few $\hat Q_m$ can be written in
terms of the field operator as $\hat Q_0=\int\ud
x\,\hat\psi^\dagger\hat\psi,$ $\int\hat Q_1=-i\int\ud
x\,\hat\psi^\dagger\p_x\hat\psi,$ and $\hat Q_2=\hat H$ is the
Hamiltonian given by Eq.~\erf{H2}.  
Unfortunately, similar second quantized expressions do not exist
for the operators $\hat Q_{m}$ for $m\ge4$ \cite{davies}. More importantly, their expectation values can be shown to diverge in almost all states other than the eigenstates of $\hat H.$ 
The reason is that their first quantized expressions contain products of Dirac deltas and higher derivatives \cite{davies,jornthesis}, and are only meaningful when evaluated on a wave function satisfying the cusp condition \erf{cusp}. Clearly, any eigenfunction of the Hamiltonian with a different coupling $c$, including the BEC wave function, will violate this condition
Note that although its expectation value is finite, even the action of the Hamiltonian is singular on such a state as it generates Dirac-$\delta$'s
\footnote{The divergence can also be verified for $N=2$ particles and quenches from the $c=0$ ground state by explicitly calculating the overlaps between the new eigenstates and initial state which is a constant. The overlaps scale as $\lam^{-2}$ for large $\lam$ which implies that $\vev{\hat Q_m}$ diverge for $m\ge4$.}. 
The diverging expectation values of the charges imply in general that the density $\rholl(\lam)$ has a $\lam^{-4}$ power-law tail instead of the usual exponential fall-off. We expect these divergences to be a generic phenomenon for interaction quenches in continuum models which has not been addressed so far. 


\paragraph*{q-boson regularization.---}

To circumvent the problem of divergences we regularize them by considering an integrable lattice regularization of the LL model, the so-called $q$-boson hopping model \cite{qboson}. The Hamiltonian is 
\begin{equation}
H_q=-\frac{1}{\delta^2} \sum_{j=1}^{M}
(B_j^{\dagger}B_{j+1}+B_{j+1}^\dagger B_j  - 2 N_j)\,,
\label{qHam}
\end{equation}
where $\delta$ is the lattice spacing of the lattice of length $M$ having periodic boundary conditions.
The operators $B_j,$ $B_j^\dagger$ and the number operator $N_j=N^\dagger_j$
satisfy the $q$-boson algebra
\be
B_j B_j^{\dagger}-q^{-2} B_j^{\dagger} B_j=1\,,\quad q>1,
\label{qalg}
\ee
with $[N_j, B_j]=-B_j$, $[N_j, B_j^\dagger]=B_j^\dagger,$ and operators at different sites commute. 
In the representation on the Fock space generated by the canonical lattice boson operators $b_j,b^\dag_j$ at each site it is possible to express the $q$-operators as
$N_j=b_j^\dagger b_j,$ $B_j= \sqrt{\frac{[N_j+1]_q}{N_j+1}}\,b_j,$ 
where
$[x]_q\equiv\frac{1-q^{-2x}}{1-q^{-2}}.$ 
Note that as $q\to1$, $[x]_q\to x$ and therefore $B^{(\dagger)}_j\to b^{(\dagger)}_j.$ The Hamiltonian is non-polynomial either in the $b$ or the $B$ operators, thus the model is interacting and the interaction is encoded in the deformation
parameter $q.$ In the naive limit $q\to 1$ we recover the system free bosons hopping on a lattice.
We are interested instead in the following {\em continuum limit}: let $\delta\to0,$ $M\to\infty,$ and
$q\to1,$ while $L$ and $c$ are kept constant:
\begin{equation}  
L=M\delta\,, \;\; c/2=\kappa\delta^{-1}\,, \text{ as }
M\to\infty \text{ and } \delta,\kappa\to 0\,, 
\label{qCont}
\end{equation}
where $\kappa$ is related to $q$ as 
$q=e^{\kappa}.$ 
Defining the continuum boson fields $\hat\psi(x=j\delta)=\delta^{-1/2}b_{j},$ 
the $q$-boson Hamiltonian \eqref{qHam} becomes the LL Hamiltonian in the limit \erf{qCont}.

The main idea behind our regularized GGE is to use the local conserved charges of the lattice model to determine the density of quasimomenta of $q$-bosons first, and to take the continuum limit yielding $\rholl(\lam)$ only as the last step. An infinite set of mutually commuting local charges can be constructed via the Quantum Inverse Scattering Method \cite{epaps}. They are of the form $I_m=\delta\sum_{j=1}^M \mathcal{J}^{(m)}_j,$
where the operators $\mathcal{J}^{(m)}_j$ act nontrivially in
$m+1$ neighboring lattice sites only. These charges are not in one-to-one correspondence with the LL operators $\hat Q_m$.

Similarly to the LL model, the common eigenstates of all $I_m$ are defined in the $N$-particle sector by
$N$ quasi-momenta $\{p_i\}$ which are solutions of the $q$-boson Bethe
equations. Under the limit \eqref{qCont} the quasi-momenta should be rescaled as $\lam_j=p_j/\delta$ in order to regain the Bethe equations of the LL model. 
In the thermodynamic limit, $N,M\to\infty,$ $\nu\equiv N/M=\text{const.}$, we introduce the quasimomentum distribution function $\rho_q(p)$. 
In terms of $\rho_q(p)$ the expectation values of the integrals of motion can be written as \cite{CHZ} 
\be
\rho_m\equiv\frac{|m| \langle I_m\rangle}{M \left(1-q^{-2|m|}\right)} =
\int_{-\pi}^{\pi}\cos{(m p)} \rho_q(p) \ud p\,,
\label{rmdef}
\ee
for $m=1,2,\dots$, and $\rho_0=\nu$, thus the expectation values are essentially the Fourier series coefficient of $ \rho_q(p).$  Here we specialized to the case where the parity symmetry is not broken and thus $\rho_q(p)$ is an even function.

\paragraph*{The density $\rholl(\lam)$.--- }

Let us evaluate now expectation values of $I_m$ in a $q$-boson state which reduces to the free boson ground state in the continuum limit, i.e. a BEC state. There is no unique choice but we pick the lattice BEC state 
$|\text{BEC} \rangle_N =  \frac{1}{\sqrt{N!}} \left(\frac{1}{\sqrt{M}}\sum_{i} b^{\dagger}_i\right)^N |0\rangle,$
where $b^{\dagger}_i$ are creation operators of {\em canonical} lattice bosons. 
Using the explicit expressions of the charge densities $\mathcal{I}^{(m)}_j$ in terms of the $B^{(\dagger)}_j$ operators, expanding these in terms of $b^{(\dagger)}_j$ 
we obtain series expansions of $\vev{I_m}$ in terms of the small parameter $\kappa$ \cite{epaps}. Combining the lowest orders of the first few $\vev{I_m}$ we confirmed that we obtain the correct value of energy in the limit, $E/L=n^3\gamma$.
We also find that 
in the continuum limit $\hat Q_4$ is divergent, as expected \cite{epaps}.

Based on the first seven charges we conjectured a pattern for the lowest orders in the expansion of the expectation values \cite{epaps}.
The distribution $\rho_q(p)$ is obtained by taking the Fourier sum, $\rho_q(p)=\frac1{2\pi}\sum_{m=-\infty}^\infty\rho_m \cos(mp)$. 
Summing up the Fourier series order by order in $\kappa$ and then taking the continuum limit we find
\be
2\pi\rholl(\lam) = \frac{n^4\gamma^2}{\lam^4}-\frac{n^6\gamma^3(\gamma-24)}{4\lam^6}+\dots\,.
\ee
The expansion of the Fourier modes $\rho_m$ in terms of $\kappa$ translates into a large momentum expansion of $\rholl(\lam)$ due to the rescaling of momenta, $\lam=p/\delta$. 
We found the expected $\lam^{-4}$ tail together with the subleading $\lam^{-6}$ tail.

To find the full $\rholl(\lam)$ function one needs a pattern for the $\rho_m$ in all orders in $\kappa$. This requires the knowledge of the expectation values of higher charges which are increasingly hard to the compute. However, for observables localized on $l$ neighboring sites the truncated GGE using the first $m\gtrsim l$ charges of size $\le m+1$ is expected to give a very good approximation 
\cite{XXZ}. Observables localized at a point in the LL model, like $g_k=\vev{:\!(\hat\psi(x)^\dagger\hat\psi(x))^k\!:}/n^k$, are the limits of operators localized on a few neighboring sites in the $q$-boson lattice system, thus we expect to capture the $g_k$ using the first few conserved $q$-boson charges. 

\begin{figure}[t]
\includegraphics[width=0.42\textwidth]{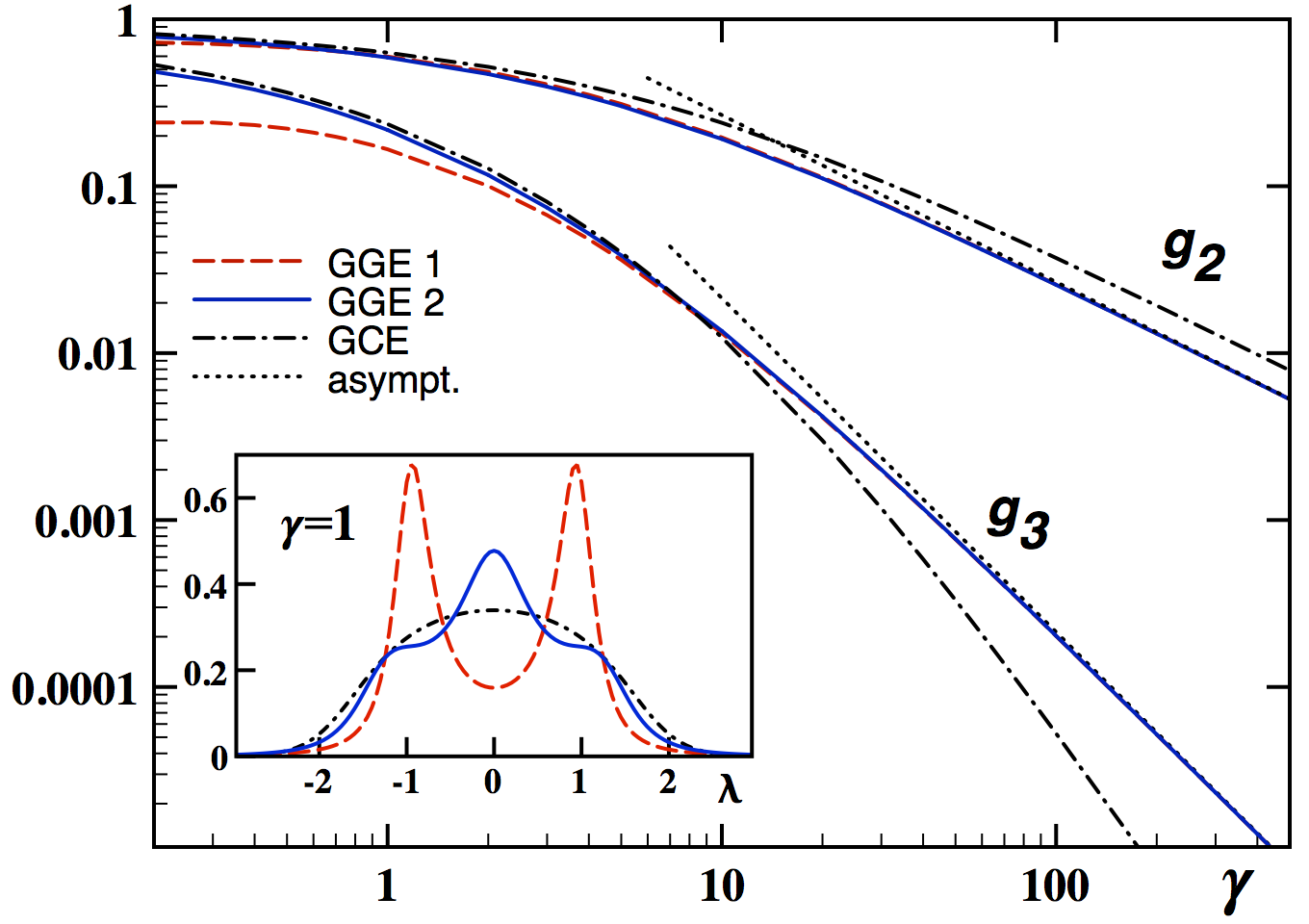}
\caption{Quench from a non-interacting initial state to
  arbitrary final interactions. {\em Main panel:} Local correlations $g_2$ and $g_3$ as
  functions of the coupling $\ga$, calculated from the two truncated Generalized
  Gibbs Ensembles (GGE) (red dashed, blue solid) and from the grand canonical ensemble (GCE)
  (dot-dashed). The asymptotic behaviors are also shown (dotted). {\em Inset:} density of quasimomenta, $\rholl^{(1)}(\lam),\rholl^{(2)}(\lam)$ in the two truncated GGE (red dashed, blue solid) and $\rholl^{\text{th}}(\lam)$ in the GCE (black dot-dashed) for $\ga=1.$}
\label{fig:g2g3}
\end{figure}

To this end, we approximate $\rho_q(p)$ by the truncated Fourier sum using the Fourier--Pad\'e approximation. Keeping charges up to $I_4$ and $I_5$, Pad\'e-approximants of different types yield the same result in the limit \cite{epaps}:
\be
\rho^{(1)}_{\text{LL}}(\lam)=\frac1{2\pi}\frac{\ga^2}{(\lam/n)^4+\ga(\ga/4-2)(\lam/n)^2+\ga^2}\,.
\label{pade1}
\ee
This result changes only when we take into account $I_6$: it becomes the ratio of a second and a sixth order polynomial in $\lam$, which we call $\rholl^{(2)}(\lam)$ \cite{epaps}. The densities are shown for $\ga=1$ in the inset of Fig.~\ref{fig:g2g3} together with the GCE density fixed by the energy and particle number only. Let us note that, interestingly, the $\ga\to\infty$ limit of both expressions gives the Lorentzian form
\be
\lim_{\ga\to\infty}\rho^{(1,2)}_{\text{LL}} (\lam)= \frac1{2\pi}\frac{4}{(\lam/n)^2+4}\,.
\label{TGlim}
\ee

\paragraph{Correlation functions in the final state.---}
Knowing the density $\rholl(\lam)$ allows us to calculate correlation functions. First we compute point-local correlators using the results of Ref.~\cite{g3} which
give analytic expressions for the local two and three-point
correlators for arbitrary states that are captured by a continuous
$\rholl(\lam)$.
We compute $g_2=\vev{:\!(\hat\psi^\dagger\hat\psi)^2\!:}/n^2$ and
$g_3=\vev{:\!(\hat\psi^\dagger\hat\psi)^3\!:}/n^3$ both for the GGE and the GCE by using the appropriate
$\rholl(\lam)$. In the latter only the energy and the particle
densities are fixed to be the same as for the GGE.
The results are shown in the main panel of Fig.~\ref{fig:g2g3}. 
The values of the correlators computed using the two Pad\'e approximants are very close 
to each other conforming with the expectation that adding more charges to the thermal GGE does not significantly change the result. 
This is an important consistency check of our truncation method. The deviations are bigger for $g_3$ which agrees with the intuition that $g_3$ is more complex than $g_2$.  The second observation is that as the difference between the two truncated results decreases for increasing $\ga$, their deviation from the GCE results $g_k^{\text{th}}$ (dotted lines) grows, the relative difference between the $g_2$ values being bigger than $20\%$ for $\ga>10$. For strong interactions the asymptotic behavior of $g_k$ can be obtained analytically. For $g_2$ we find $g_2\sim8/(3\ga)$ and $g_2^{\text{th}}\sim4/\ga$ implying a factor of $3/2$ between the two. For $g_3$ even the power laws are different: $g_3\sim32/(15\ga^2)$ while $g_3^{\text{th}}\sim72/\ga^3$.

\paragraph*{Strongly interacting final state.---}

For large coupling the system is in the fermionized TG
regime since the strong repulsion induces an effective Pauli principle in real space. In the special case of the quench from $c=0$ to $c=\infty$ the overlaps between the initial state and the final TG eigenstates are explicitly known \cite{gritsev}. Only states defined by a set of $\{\lam_i,-\lam_i\}$ pairs have nonzero overlaps which are $\langle{\lam_i}|\text{BEC}\rangle\propto 1/\prod_{\lam_i>0}\lam_i$. The overlaps are the necessary ingredients in the formalism of Ref.~\cite{js-fabian} to compute the saddle point density. Solving the generalized TBA equations
we obtain the simple result $2\pi\rholl(\lam)=1/(1+\lam^2n^2/4)$ (see also Ref.\ \cite{j-s}) which exactly matches the $\ga\to\infty$ limit of our Pad\'e-approximants, Eq.~\erf{TGlim}. The fact that the two derivations are completely independent gives a strong evidence for the correctness of the result.

\begin{figure}[t]
\includegraphics[width=0.42\textwidth]{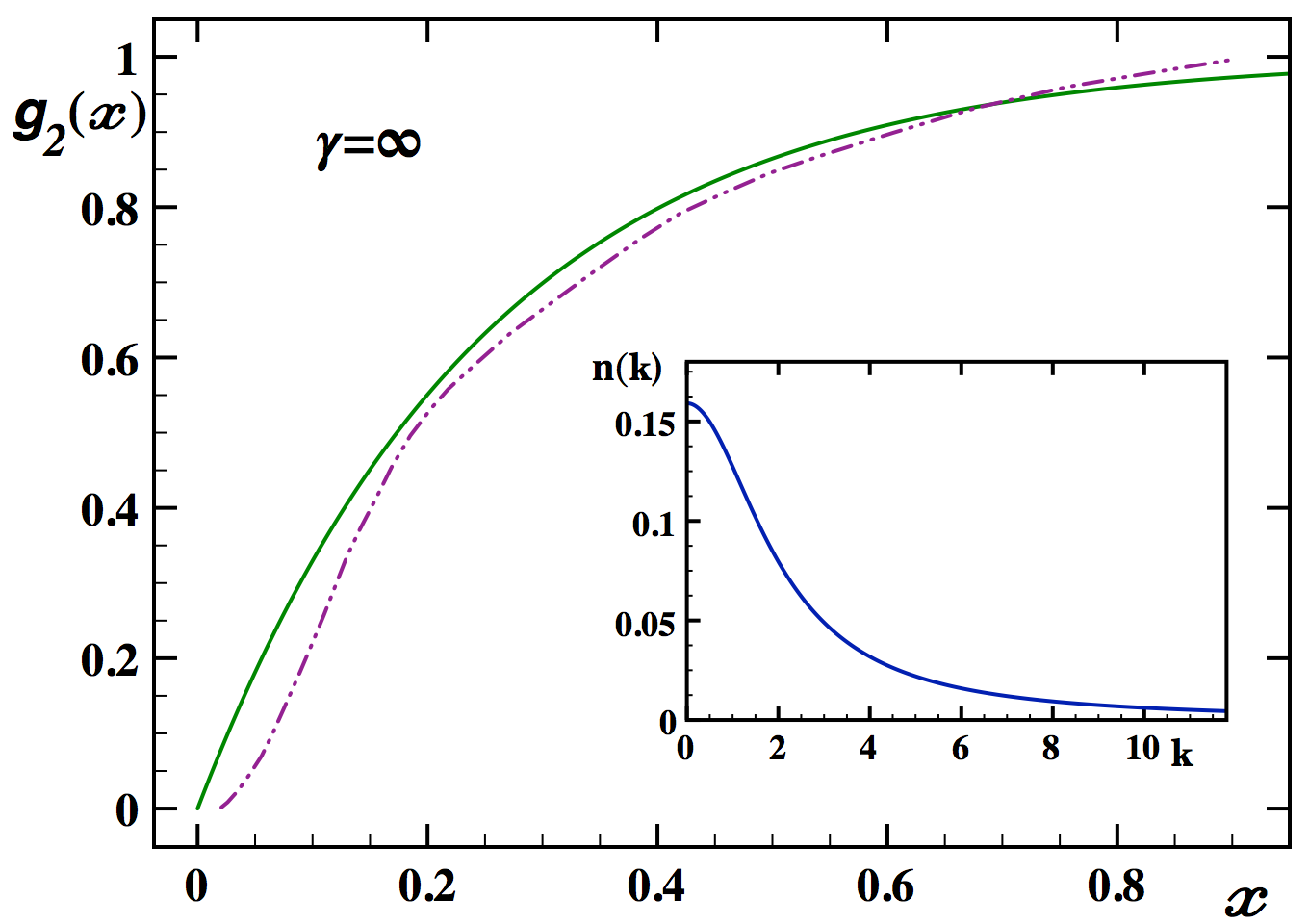}
\caption{Quench to the TG regime ($\ga=\infty$). {\em Main panel:} Equal time density-density correlation function. We compare GGE/saddle point (green solid) values with the large time result of a numerical solution of the dynamics of Ref.~\cite{gritsev}(purple dot-dot-dashed). {\em Inset:} Momentum distribution function.}
\label{fig:gx}
\end{figure}

Bosonic correlation functions can now be calculated by first fermionizing the field operators 
using Jordan--Wigner strings,
$\hat\psi(x) = 
\exp[i\pi\int_{-\infty}^x\hat\psi^\dagger_\text{F}(z)
\hat\psi_\text{F}(z)dz]\hat\psi_\text{F}(x)$,
and then exploiting free fermionic correlators of
$\hat\psi_\text{F}$. Let us consider the equal time correlation
$G(x) = \langle \hat\psi^\dagger(x)\hat\psi(0)\rangle$
in the saddle point distribution of Eq.~\erf{TGlim}. After introducing a
lattice discretization, the long chain of operators is amenable to a Wick expansion
using as a building block the {\em fermionic} two point function. Since for $\ga=\infty$ the quasimomenta coincide with the physical momenta, this is given
by the Fourier transform of the density \erf{TGlim}, $ G_{\rm
  FF}(x) =\int \ud\lam \rho_{\text{s}}(\lam) e^{i\lam x} =
e^{-2n|x|}.$ The Wick expansion can be
recast as a Fredholm-like determinant \cite{ripplezvon}
that finally leads to $G(x) =e^{-2n|x|}$. This simple result is drastically different from the corresponding GCE result, $G_{\text{th}}(x)\stackrel{\ga\to\infty}{\longrightarrow} e^{-\ga (nx)^2/2}$, which approaches an infinitely narrow Dirac-$\delta$ in the TG limit. Since $G(x)=G_\text{FF}(x)$, the experimentally accessible bosonic momentum distribution, $n_\text{B}(k)$, is thus equal to $\rholl(k)$ given by Eq.~\erf{TGlim}, plotted in the inset of Fig.~\ref{fig:gx}.

We can also compute the density-density correlation function
$
g_2(x)=\vev{\hat\psid(x)\hat\psid(0)\hat\psi(0)\hat\psi(x)}/n^2
$
for large final $\ga$ using the first few terms of the
infinite series given in Ref.~\cite{bogoliubov}. 
In the large $\ga$ limit the leading order for
arbitrary $\rho(\lam)$ is given by 
$
g_2(x)\approx 1-\left(\int\ud\lam\,\rho(\lam)e^{i\lam x}\right)^2.
$ 
Using $\rho_s(\lam)$ we obtain $g_2(x)= 1-e^{-4n|x|}$,
which agrees very well with the large time result of the numerical solution of the time evolution in Ref.~\cite{gritsev} based on the exact overlaps (see main panel of Fig.~\ref{fig:gx}). To the best of our knowledge this is one of the first demonstrations in a continuum integrable model that the GGE value of an observable agrees with its actual large time asymptotics.

\paragraph{Summary.---}

Extending the studies of the post-quench behavior of many-body systems to a non-quadratic continuum model, we investigated the large time behavior of the Lieb--Liniger model after an interaction quench using analytic techniques by combining the Generalized Gibbs Ensemble and Bethe Ansatz integrability of the model and its lattice discretization. We pointed out the divergence of local charges in the initial state that prevents the naive application of the GGE methodology. We expect this to be a generic phenomenon for interaction quenches in continuum models which deserves further study. For a non-interacting initial state and arbitrary final interactions, we evaluated local correlations and found deviations from the thermal predictions. These are experimentally accessible through the measurement of the photoassociation rate ($g_2$) and the inelastic three-body loss ($g_3$) in cold atom experiments. We computed two-point correlation functions exactly for quenches to the femionized Tonks--Girardeau regime and found excellent agreement with a recent numerical simulation of the time evolution.

\paragraph*{Note added.---} During the completion of the manuscript two preprints appeared \cite{XXZ} which considered the truncated GGE in the BA solvable XXZ spin chain.

\paragraph*{Acknowledgments.---} 

This work was initiated by Adilet Imambekov and has greatly benefited from his ideas and his detailed calculations. Adilet tragically passed away before the completion of the work, but we will always keep him in our memories as a wonderful person, a great mentor and an excellent scientist.

We are grateful for enlightening discussions with Pasquale Calabrese, Jacopo De Nardis, Michael Brockmann, Bram Wouters, Spyros Sotiriadis, Bal\'azs Pozsgay, G\'abor Tak\'acs.
We acknowledge funding from The Welch Foundation, Grant No. C-1739,
from the Sloan Foundation and from the NSF Career Award No. DMR-1049082. M. K. acknowledges ERC for financial support under Starting Grant 279391 EDEQS. J.-S. C. acknowledges support from FOM and the NWO of the Netherlands.


\clearpage
\newpage

\clearpage 
\setcounter{equation}{0}%
\renewcommand{\theequation}{S\arabic{equation}}
\setcounter{page}{1}

\onecolumngrid

\begin{center}
{\Large Supplementary Material for EPAPS \\ 
Interaction quenches in the Lieb--Liniger model }
\end{center}

\section{Local conserved charges in the $q$-boson hopping model}

Integrals of motion of the $q$-boson hopping
model can be constructed using the Quantum Inverse Scattering
Method. The $L$-operator for the model is given by
\begin{equation}
L_{j}(\lambda)= \left(
\begin{array}{cc}
e^{\lambda} & \chi B_j^\dagger \\
\chi B_j & e^{-\lambda}
\end{array}
\right)\,, \label{qL}
\end{equation}
where $\chi=\sqrt{1-q^{-2}}=\sqrt{1-e^{-2\kappa}}.$
The monodromy matrix $T(\lambda)$ is defined as a matrix product of
the $L$-operators over all the lattice sites
\be
T(\lambda)=L_M(\lambda)L_{M-1}(\lambda)\cdots L_1(\lambda)\,,
\label{Tdef}
\ee
and the transfer matrix $\tau(\lambda)$ is given by the trace over the
matrix space of the monodromy matrix
\begin{equation}
\tau(\lambda)=\mathrm{Tr}\, T(\lambda)\,.
\label{taudef}
\end{equation}
For any $\lam$ and $\mu$ the transfer matrices commute: $[\tau(\lam), \tau(\mu)]=0$, which implies that $\tau(\lam)$ is a generating function of the
conserved charges. Many different sets can be generated since any analytic
function of $\tau(\lambda)$ can play the role of the generating
function. We consider the set consisting of {\em local} charges that can be 
written in the form
\begin{equation}
I_m=\delta\sum_{j=1}^M \mathcal{J}^{(m)}_j\,, 
\label{IJ}
\end{equation}
where the operators $\mathcal{J}^{(m)}_j$ act nontrivially in
$m+1$ neighboring lattice sites only.  This set is obtained by the formula 
\be
I_{m}= \left.\frac{1}{(2m)!} \frac{d^{2m}}{d \zeta^{2m}}
\ln\left[\zeta^M\tau(\zeta)\right] \right|_{\zeta\to 0}\,, \qquad
m=1,2,3,\ldots\,.
\ee
where we introduced the variable $\zeta=e^{\lam}$. The local operators $\mathcal{J}^{(1)}(n),$
$\mathcal{J}^{(2)}(n)$ and $\mathcal{J}^{(3)}(n)$ are
\begin{align}
\label{J1}
&\mathcal{J}^{(1)}(n)=\frac{1}{\delta}\chi^2 B^\dagger_j B_{j+1}\,, \\
\label{J2} &\mathcal{J}^{(2)}(n)=
\frac{1}{\delta}\chi^2\left(1-\frac{\chi^2}{2}\right) \left(
B^\dagger_j B_{j+2} - \frac{\chi^2}{2-\chi^2} B_j^\dagger
B_j^\dagger B_{j+1}B_{j+1}- \chi^2 B_j^\dagger B_{j+1}^\dagger
B_{j+1}B_{j+2} \right)\,,
\end{align}
and
\begin{multline}
\mathcal{J}^{(3)}(n)= \frac1\delta\chi^2
\left(1-\chi^2+\frac{\chi^4}{3} \right)
\left(\vphantom{\frac{\chi^4}{3}} B^\dagger_j B_{j+3} -\chi^2
B^\dagger_jB^\dagger_j B_{j+1}B_{j+2}-
 \chi^2 B^\dagger_j B^\dagger_{j+1} B_{j+1}B_{j+3} \right.\\
- \chi^2 B^\dagger_j B^\dagger_{j+1} B_{j+2}B_{j+2}- \chi^2
B^\dagger_j B^\dagger_{j+2} B_{j+2}B_{j+3} + \frac{\chi^4}{3-3
\chi^2+\chi^4}B^\dagger_j B^\dagger_j B^\dagger_j
B_{j+1}B_{j+1}B_{j+1}
\\
+\chi^4 B^\dagger_j B^\dagger_j B^\dagger_{j+1}
B_{j+1}B_{j+1}B_{j+2}+ \chi^4 B^\dagger_j B^\dagger_{j+1}
B^\dagger_{j+1} B_{j+1}B_{j+2}B_{j+2}
\\
+ \left. \chi^4 B^\dagger_j B^\dagger_{j+1} B^\dagger_{j+2}
B_{j+1}B_{j+2}B_{j+3} \vphantom{\frac{\chi^4}{3}} \right) \label{J3}\,.
\end{multline}

The integrals $I_m$ are not Hermitian operators. Using the involution $[\tau(\zeta)]^\dagger=\tau(\zeta^{-1})$ it can be shown that $[I^{\dagger}_m, I_n]=0$ for any $m, n$. As the number operator $N=\sum_j N_j=\sum_j b_j^\dagger b_j,$ is non-polynomial in the $B^{(\dagger)}_j$ operators while the charges $I_m$ are, it cannot be expressed as a finite linear combination of
the $I_m.$ However, $N$ commutes with any monomial containing an equal number of
the creation and annihilation operators thus $[N,I_m]=0.$
It is convenient to use the notation
$N\equiv I_0.$ 
The Hamiltonian \eqref{qHam} can then be written as
\begin{equation}
H_q=-\frac1{\chi^2\delta^2}(I_1+I_{-1}-2\chi^2 I_0)\,. 
\label{HviaI}
\end{equation}

\section{Expectation values of the charges in the initial state}

We need to evaluate the expectation values of local charges $I_m$ in a state which transforms into the continuum BEC-state in the limit \erf{qCont}. We pick here the state
\be
|\text{BEC} \rangle_N =  \frac{1}{\sqrt{N!}} \left(\frac{1}{\sqrt{M}}\sum_{i} b^{\dagger}_i\right)^N |0\rangle\,,
\label{bec2}
\ee
since it has the nice property (established by commuting annihilation operators one by one)
\be
b^\alpha_j  |\text{BEC} \rangle_N = \sqrt{\frac{N}{M}}\cdots \sqrt{\frac{N-\alpha+1}{M}}|\text{BEC} \rangle_{N-\alpha} \approx  \nu^{\alpha/2}|\text{BEC} \rangle_{ N-\alpha}\,,
\label{belem}
\ee
where the approximate relation is valid in the thermodynamic limit (TDL) when we are interested in $\alpha$ that does not scale proportionally to the system size. Note that  as long as we are interested in evaluation of expectation values of {\it normal ordered}  operators over BEC state in the TDL, one can also use the coherent state form of the BEC
\be
|\text{BEC},c\rangle = \prod_j e^{-\nu/2 + \sqrt{\nu} b^\dagger_j} |0\rangle\,, 
\label{BEC_c}
\ee
which has the same matrix elements as state (\ref{belem}).

In what follows, we compute expectation values of the local charges by computing first the building blocks, on-site monomials, based on expanding $B^{(\dagger)}_i$ in terms of $b^{(\dagger)}_i$ and normal ordering. For most of the matrix elements we can only derive expansions in powers of $\kappa$ (but not making any assumptions about $\nu$). We will start from
\be
B_j=b_j\sqrt{\frac{[N+1]_q}{N+1}}\approx b_j\left(1-\frac{\kappa}{2}N_j+\frac{\kappa^2}{24}N_j(5N_j+4)+\dots\right) =
b_j-\frac{\kappa}{2}b_j^{\dagger}b_jb_j+\frac{\kappa^2}{24}(5b_j^{\dagger}b_j^{\dagger}b_jb_jb_j+9b_j^{\dagger}b_jb_j)+\dots \,.
\ee
The evaluation of its expectation value in the state \erf{BEC_c} leads to
\be
\langle \text{BEC},c|B_j|\text{BEC},c\rangle = \sqrt{\nu} - \frac12 \kappa \nu^{3/2} + \frac{ \kappa^2}{24}\left(9\nu^{3/2} +5\nu^{5/2}\right)+\dots\,.
\ee
In a similar way we obtain
\be
\langle \text{BEC}|B_j^{\dagger} B_j|\text{BEC}\rangle  =
\nu - \kappa \nu ^2 + \kappa^2 \nu^2 -\frac23\nu^2\kappa^3+\frac23\nu^3\kappa^2
+\dots\,.  
\ee
We note that for this combination a closed form expression exists, $\langle \text{BEC}|B_j^{\dagger} B_j|\text{BEC}\rangle  = (1-e^{-\left(1-q^{-2}\right)\nu})/(1-q^{-2}).$ These and similar on-site matrix elements are the only type needed to systematically evaluate the expectation values of any polynomial of $B^{(\dagger)}$ operators acting on different sites over the BEC. Indeed, due to the factorization of the wave function on different sites in the coherent state representation (\ref{BEC_c}) one can treat different sites separately. 

Let us now use these matrix elements to evaluate the first $\rho_m,$ $m=1,\dots,6.$ From Eqs.~(\ref{J1},\ref{J2}) and from the definition (\ref{rmdef}) we have
\be
\rho_1=\frac{1}{M}\sum_j \langle B^\dagger_j B_{j+1} \rangle\,,\quad
\rho_2=\frac{1}{M} \sum_j  \left\langle 
B^\dagger_j B_{j+2} - \frac{\chi^2}{2-\chi^2} B_j^\dagger
B_j^\dagger B_{j+1}B_{j+1}- \chi^2 B_j^\dagger B_{j+1}^\dagger
B_{j+1}B_{j+2} \right\rangle\,,\;\text{etc.}
\ee
Due to translational invariance we need to evaluate the expectation value only for a single value of $j.$
We find
\begin{subequations}
\label{rhoseries}
\begin{align}
\rho_1&=\nu - \frac12\ga \nu^3 + \frac3{16} \ga^2 \nu^4 + \left(\frac{1}6\ga^2 - \frac{7}{192}\ga^3\right) \nu^5 - \frac{11}{96}\ga^3 \nu^6 +\dots\,,\\
\rho_2&=\nu - 2 \ga \nu^3 + \frac{15}{16}\ga^2 \nu^4 + \left(\frac53 \ga^2 - \frac{55}{192} \ga^3\right)\nu^5 
 + \left(-\frac{51}{32} \ga^3 + \frac1{16}\ga^4\right) \nu^6+\dots\,,\\
\rho_3&=\nu - \frac92 \ga \nu^3 + \frac{43}{16}\ga^2 \nu^4 + \left(\frac{15}2 \ga^2 - \frac{73}{64} \ga^3\right) \nu^5 + \left(-\frac{859}{96} \ga^3 + \frac{73}{192} \ga^4\right) \nu^6 +\dots\,,\\
\rho_4&=\nu - 8 \ga \nu^3 + \frac{95}{16} \ga^2 \nu^4 + \left(\frac{68}3 \ga^2 - \frac{619}{192} \ga^3\right) \nu^5 + \left(-\frac{3137}{96} \ga^3 + \frac{269}{192} \ga^4\right) \nu^6+\dots\,,\\
\rho_5&=\nu - \frac{25}2 \ga \nu^3+ \frac{179}{16} \ga^2 \nu^4 + \left(\frac{325}6 \ga^2 - \frac{1423}{192} \ga^3\right) \nu^5 + \left(-\frac{2953}{32} \ga^3 + \frac{379}{96} \ga^4\right) \nu^6 + \dots\,,\\
\rho_6&=\nu - 18 \ga \nu^3 + \frac{303}{16} \ga^2 \nu^4 + \left(111 \ga^2 - \frac{949}{64} \ga^3\right) \nu^5 + \left(-\frac{21049}{96} \ga^3 + \frac{299}{32} \ga^4\right) \nu^6+\dots\,,
\end{align}
\end{subequations}
where we use $\nu$ as small parameter by the relation 
\be
\kappa=\ga\nu/2\,. 
\label{kappagamma}
\ee

Naively, combining various $\rho_m$ and taking the limit one can obtain moments of the $\rholl(\lam)$ density, i.e. the expectation values of the charges $\hat Q_m$ of the LL model. However, this must be done with care. First, the limits of integration are strictly speaking not $\pm \infty$,  but $\pm \pi/\delta,$ which matters if the LL moments are divergent (as expected). Second, the scaling limit \erf{qCont}, Eq.~ \erf{kappagamma} as well as the relations $\lam=p/\delta$ and $\rholl(\lam)=\rho_q(\delta\,\lam)$ may have higher order corrections which would mix the orders.

In spite of the problems mentioned above, the energy can be obtained if the $\rho_{\text{LL}}(\lam)$ has at most a $\lam^{-4}$ tail:
\be
-\frac14(\rho_1+\rho_{-1}-2\rho_0) = 
\int_{-\pi}^\pi\ud p \,(p^2-\frac16p^4+\dots)\rho_q(p)=
\int_{-\pi/\delta+\dots}^{\pi/\delta+\dots}\ud \lam(\delta+\dots) (\delta^2\lam^2-\frac16\lam^4\delta^4+\dots)(\rho_{\text{LL}}(\lam)+\dots)\,,
\ee
where the dots stand for higher order terms in $\delta$. The first parenthesis comes from the unknown higher order terms of the relation $p=\delta\cdot\lam+\dots$, this also generates terms in the middle parenthesis. Now let us make the {\em assumption} that this relation, as well as the relation between $\rho(p)$ and $\rho_{\text{LL}}(\lam)$ does not have higher powers of $\lambda$. Under this assumption each power $\lam^{2n}$ comes with at least $\delta^{2n+1}$ in the integrand which implies that although the integrals of higher powers seem to diverge, with the $\delta$-powers in their coefficients all of them scale as $\delta^4$, while the quadratic term scales as $\delta^3$. Thus it is safe to take the $\delta\to0$ limit after dividing by $\delta^3$ and we are left with
\be
-\frac1{2\delta^3}(\rho_1-\rho_0) \to \int_{-\infty}^\infty\ud \lam\,\rho_{\text{LL}}(\lam)\lam^2\,.
\ee
Since $\lim_{\delta\to0}\left[-\frac1{2\delta^3}(\rho_1-\rho_0)\right]=n^3\gamma$, the energy density is correctly reproduced, as expected.

In a similar fashion, one can formulate a condition on whether the $2n$-th moment of the $\rholl(\lam)$ distribution is divergent. For this one again needs to pick the right combination 
of $\rho_m$ with $m\leq n.$ In particular, $Q_4$ is divergent if 
\be
\rho_2+\rho_{-2}-2\rho_0-4 (\rho_1+\rho_{-1}-2\rho_0) \propto \delta^4\,, 
\label{Q4crit}
\ee
as opposed to $\propto \delta^5$. From the expansion in Eqs.~\erf{rhoseries} we find
\be
\rho_2-\rho_0-4 (\rho_1-\rho_0) = \frac3{\ga^2}\kappa^4+\dots\,,
\ee
thus $\int\ud\lam \rholl(\lam)\lam^4=\vev{Q_4}/L$ is divergent.


\section{Pattern for expectation values in the BEC state}
\label{pattern}
Based on the Taylor expansions in Eqs.~\erf{rhoseries} one can find a pattern for the coefficients of the different orders. They turn out to be low order polynomials in $m$:
\begin{multline}
\rho_m=\nu-\frac{m^2}2\gamma\nu^3+\frac{m^3+2m-\frac34}{12}\gamma^2\nu^4+\left(\frac{m^2(m^2+1)}{12}\gamma^2-\frac{m^4+4m^2-3m+\frac32}{96}\gamma^3\right)\nu^5\\
+\left(-\frac{m^5+5m^3-\frac52m^2+\frac23m+\frac5{12}}{40}\gamma^3+\frac{m^5+\frac{20}3m^3-\frac{15}2m^2+\frac{29}6m-5}{960}\gamma^4\right)\nu^6+\mathcal{O}(\nu^7)\,.
\label{rhom}
\end{multline}
%
The reasonably simple rational coefficients and their structure provide strong evidence that  the polynomial dependence on $m$ is correct. The order of the coefficient polynomial of $\nu^k$ is $k-1$ and, interestingly, the subleading orders ($m^{k-2}$) are always missing. As we will show now, the first property is necessary in order to have a finite scaling limit of the $\rho(p)$ function, i.e. a finite $\rho_{\text{LL}}(\lam)$. 

The $\rho(p)$ distribution function is the Fourier sum of the $\rho_m$. It is clear that the scaling limit and this Fourier transformation do not commute: if we take the limit before computing the sum we get $\rho_m\equiv0$. For the computation of the Fourier sum order by order in $\nu$ one needs to calculate the building blocks
\begin{align}
\sum_{m=-\infty}^\infty m^{2l} \cos(mp)&=0\,,\\
\sum_{m=-\infty}^\infty m^{2l-1} \cos(mp)&=\frac{\sum_{j=0}^{l-1}c_j\cos(jp)}{\sin^{2l}\left(\frac{p}2\right)}
\longrightarrow \frac{2^{2l}\sum_{j=0}^{l-1}c_j}{\delta^{2l}\lam^{2l}}\,,
\end{align}
where the $c_j$ are real numbers. This must be multiplied by $\delta^{2l}$ to be neither divergent nor zero. Thus the fact that in Eq.~\eqref{rhom} the highest power of $m$ in the coefficient of $\nu^{k}$ is $k-1$ implies that $\rho_{\text{LL}}(\lam)$ is finite. Moreover, only the highest powers of $m$ in the coefficient polynomials of the even orders of $\nu$ contributes. This is important, because we know the relation $\kappa=\gamma\nu/2$ only to leading order. Adding potential sub-leading terms, $\kappa=\gamma\nu/2+a_1\nu^2+a_2\nu^3+\dots$, generates terms in each order of $\nu$ which however have a sub-leading $m$-dependence, thus they will do not affect the result in the continuum limit.

Taking the Fourier sum we obtain
\begin{multline}
2\pi \rho(p)=\rho_0+2\sum_{m=1}^\infty \rho_m \cos(mp) = \nu + 2\sum_{m=1}^\infty\left(\frac1{12}\gamma^2\nu^4 m^3+\frac{\gamma^3(\gamma-24)}{960}\nu^6 m^5\right)+\text{[``subleading terms'']}+\dots  \\
=\nu+2\left(\nu^4\frac{\gamma^2}{12}\,\frac{2+\cos(p)}{8\sin^4(p/2)}-
\nu^6\frac{\gamma^3(\gamma-24)}{960}\,\frac{33+26\cos(p)+\cos(2p)}{32\sin^6(p/2)}+\dots\right)\,.
\end{multline}
Taking the continuum limit \erf{qCont} together with $p=\delta \lam$ we find
\be
\rho_{\text{LL}}(\lam) = \frac{n^4\gamma^2}{\lam^4}-\frac{n^6\gamma^3(\gamma-24)}{4\lam^6}+\dots\,.
\label{tails}
\ee
We see that the expansion of the Fourier modes $\rho_m$ in terms of $\delta$ or $\nu$ is equivalent to a {\em large momentum expansion} of the LL density of roots $\rho_{\text{LL}}(\lam)$. We did find the expected $\lam^{-4}$ tail together with the subleading $\lam^{-6}$ tail. 

Observe that going to higher charges and to higher powers in $\nu$ go side by side: if one only expands the $\rho_m$ up to a fixed order in $\nu$ then one does not gain anything from considering many more charges because the polynomial pattern found from the lower ones already determines them. Conversely, having only a few charges does not allow one to determine the high order polynomial coefficients of the higher orders of $\nu$.

A key step in all the above is the rescaling of momenta, $\lam=p/\delta$. This is how lower orders of $\nu$ may eventually disappear and arbitrary high powers of $\nu$ may survive in the limit. Consequently, the large momentum expansion structure can be heuristically understood by realizing that we need to resolve the vicinity of $p=0$ very well, because this region will be blown up to be the entire domain in $\lam$. Thus it is not very surprising that many Fourier modes are necessary and one needs to know them very precisely. Any truncation or approximation affects the small $\lam$ region, so perturbatively we approach from large $\lam$.

\section{Pad\'e--Fourier approximation}

Let us consider the truncated Fourier sum,
\be
\rho^{[l]}(p)=\rho_0+2\sum_{m=1}^l \rho_m \cos(mp) = 
\left(\frac{\rho_0}2+\sum_{m=1}^l \rho_m e^{imp}\right) +\{p\to-p\}=
\left(\frac{\rho_0}2+\sum_{m=1}^l \rho_m z^m\right) +\{z\to1/z\}\,,
\ee
where we introduced $z=e^{ip}$. The parenthesis is a truncated Taylor expansion to which we apply the Hermite--Pad\'e approximation technique: we find a rational function of $z$ such that the first $l$ terms in its Tayor expansion matches our truncated expansion. The $(n,m)$-type Pad\'e-approximant is a ratio of an $n$th order and an $m$th order polynomial ($n+m=l$). We reintroduce the variable $p$ in the approximants and then we take the continuum limit. The $(2,2)$, $(3,2)$, $(2,3)$, $(4,2)$ and $(2,4)$ Pad\'e-approximants all give the same result, Eq.~\erf{pade1}:
\be
\rho^{(1)}_{\text{LL}}(\lam)=\frac1{2\pi}\frac{\ga^2}{(\lam/n)^4+\ga(\ga/4-2)(\lam/n)^2+\ga^2}\,.
\ee
Comparing with Eq.~\erf{tails} this has the correct $\lam^{-4}$ tail but not the $\lam^{-6}$ one. The latter is reproduced by the Pad\'e-approximant of type $(3,3)$:
\be
\rho^{(2)}_{\text{LL}}(\lam)=\frac1{2\pi}\frac{4 \ga^2 \left(\bar\lam^2 + \ga (\ga+2)\right)}{(4 \bar\lam^2 + \ga^2)\left(\bar\lam^4 +  (\ga-4) \ga\,\bar\lam^2 + 4 \ga^2\right)}\,,
\ee
where $\bar\lam=\lam/n$. The $\ga\to\infty$ limit of both $\rho^{(1)}(\lam)$ and $\rho^{(2)}(\lam)$ is given by Eq.~\erf{TGlim}.

\end{document}